\documentclass[aps,prd,groupedaddress,showpacs,showkeys]{revtex4}
\usepackage{epsfig}
\newcommand{\be}[1]{\begin{equation} \label{(#1)}}
\newcommand{\eq}{\begin{eqnarray}}
\newcommand{\ee}{\end{equation}}
\newcommand{\en}{\end{eqnarray}}
\newcommand{\ba}[1]{\begin{eqnarray} \label{(#1)}}
\newcommand{\ea}{\end{eqnarray}}

\newcommand{\rf}[1]{(\ref{(#1)})}

\def \gsim {\mbox{${}^> \hspace*{-7pt} _\sim$}}

\begin{document}

\title{On Sterile neutrino explanation of LSND and MiniBooNE anomalies}
\author{ Claudio Dib, Juan Carlos Helo,  Sergey Kovalenko, Ivan Schmidt  \vspace*{0.3\baselineskip}}
\affiliation{Universidad T\'ecnica Federico Santa Mar\'\i a, \\
Centro-Cient\'\i fico-Tecnol\'{o}gico de Valpara\'\i so, \\
Casilla 110-V, Valpara\'\i so,  Chile \vspace*{0.3\baselineskip}\\}

\date{\today}

\begin{abstract}
We examine the compatibility between existing experimental data and a recently proposed explanation of the LSND and
MiniBooNE anomalies, given in terms of a sterile neutrino $N$ whose decay is dominated by a radiative mode.
We find that current experimental data on
$\tau\rightarrow \mu\nu\nu\gamma$ decays are compatible with the sterile neutrino parameters required for the explanation of the anomalies, but
$K\rightarrow \mu\nu\gamma$ shows a marginal tension with those parameters. We also propose experimental cuts on radiative $K$ decays that could test the sterile neutrino hypothesis better. Finally, we study the contribution of this sterile neutrino to  $K\to\mu\nu e e$, and find that measurements of this process would provide powerful tests for the sterile neutrino explanation of the LSND and MiniBooNE anomalies, if the experimental cut on the invariant mass of the $e^{+}e^{-}$ pair could be reduced from its current value of 145  MeV to a value below 40 MeV.
\end{abstract}




\maketitle

\newpage

Neutrino oscillation experiments have proven that neutrinos are massive, although very light particles, and that they  exhibit flavor mixing. In order to give mass to neutrinos,  most models introduce sterile (or right-handed) neutrinos, which generate the masses of the ordinary neutrinos via a see-saw mechanism  or its modifications \cite{see-saw}, \cite{Valle}.
This mechanism gives masses to the three light neutrinos, leaving open  the possibility of having one or more additional heavy neutrinos $N$, which would be sterile with respect to electroweak gauge interactions. If this is the case, the sterile neutrinos $N$  in general will contain a certain admixture of the active flavors
$\nu_{e,\mu,\tau}$, parametrized by the corresponding elements of a neutrino mixing matrix
$U_{eN}, U_{\mu N}, U_{\tau N}$.
Therefore, $N$  can participate in charged and neutral current interactions of the Standard Model (SM), contributing to various  processes. If a sterile neutrino with mass $m_N \lesssim 100$ MeV is produced in an intermediate state, it  would typically decay into three leptons, but a radiative decay is also possible if a nonzero transition magnetic moment $(\mu_{tr})$ between the $N$ and $\nu$ mass states is introduced \cite{Gninenko}-\cite{Vogel}. Usually the radiative decay of the sterile neutrino is assumed to be negligible compared to its decay into three leptons. However, it has been recently proposed that a sterile neutrino $N$ with a dominant radiative decay mode $N\rightarrow \nu \gamma$ and with mass $m_{N}$, mixing strength $U_{\mu N}$  and lifetime  $\tau_{N}$ in the range \cite{Gninenko,LimitGninenko}
\ba{Range}
40\textrm{ MeV} \lesssim m_N \lesssim 80\textrm{ MeV}  \ \  \  , \ \ \
10^{-3} \lesssim |U_{\mu N}|^2 \lesssim 10^{-2}   \ \  \  , \ \ \
 \tau_N \lesssim 10^{-9} \textrm{ s} ,
\ea
may be the source of the LSND \cite{LSND} and MiniBooNE \cite{MiniBooNE} experimental anomalies. In order to search for this sterile neutrino in an independent way,   a  new muon decay experiment  \cite{Gninenko:2011xa}, direct searches through  $K$ meson decays \cite{LimitGninenko} and searches at neutrino telescopes \cite{Experiment2} have already  been proposed. It was also shown that the sterile neutrino parameters with the values in the range \rf{Range} are in some tension with the radiative muon capture \cite{McKeen:2010rx}. However, this tension can be relaxed \cite{Gninenko:2011xa} and does not have an impact on the region \rf{Range}. Other constraints relevant for the range \rf{Range} have been derived in \cite{Kusenko:2004qc} from the accelerator and Super-Kamiokande results.

Here  we consider the restrictions on the sterile neutrino $N$ parameters that can be deduced from the existing experimental data on radiative $K$-meson and $\tau$-lepton decays. The purpose of this note is to check whether  these restrictions are consistent or exclude some of the values in Eq.\  \rf{Range}, necessary for the explanation of the MiniBooNE and LSND anomalies.
Specifically, we analyze the contribution of the sterile neutrino $N$  to the following decays:
\begin{eqnarray}\label{DECAYS}
 K^+ \rightarrow \mu^{+} \nu \gamma  \ \  \  , \ \ \
 \tau^{-} \rightarrow \mu^-  \nu  \nu \gamma  .
\end{eqnarray}
Here  $\nu$ denotes the standard light neutrino or antineutrino, dominated by any of the neutrino flavors $\nu_{e}, \nu_{\mu}, \nu_{\tau}$. These decays receive their known SM contributions, which alone give good agreement with the experimental data.  However, they also proceed
according to the diagrams shown in Figs.\ \ref{fig-1}(a) and  1(b),  with the
sterile neutrino $N$ as an intermediate particle. When  $N$  is off-shell, the contribution of these diagrams
is negligibly  small  \cite{K-decPaper,tau-decPaper} and far from experimental reach. On the other hand, there exist specific domains of sterile neutrino masses $m_{N}$ where $N$ comes close to its mass-shell,
leading to an enormous resonant enhancement  \cite{K-decPaper, tau-decPaper} of the diagrams in Fig.\ref{fig-1}.  These domains, for the $K$ and $\tau$ decays in Eq.\   (\ref{DECAYS}) are respectively:
\ba{Domain}
m_N < m_K - m_\mu  \ \  \  , \ \ \
m_N < m_\tau - m_\mu  ,
\ea
where the light neutrino mass,  $m_\nu$, has been neglected. The mass domains in Eq.\ \rf{Domain}
cover completely the sterile neutrino mass range of Eq.\  \rf{Range}, proposed for the explanation of the LSND and the MiniBooNe anomalies.  Therefore, if there is a neutrino $N$ with a mass which is appropriate to explain the anomalies, then the $K$ and $\tau$ radiative decays will necessarily have a contribution from this neutrino $N$ close its mass-shell. This means that an intermediate sterile neutrino is produced at the corresponding vertex on the left of the diagrams in Fig.~\ref{fig-1}, propagates as a free unstable particle, and then decays at the corresponding vertex on the right.  Accordingly, the decay rate formulas for the reactions  $K,\ \tau  \rightarrow X \nu \gamma$ can be represented in the narrow width approximation ($\tau_N^{-1}\ll m_{N}$) as the product of two factors:  the $K$ or $\tau$ decay rate into the sterile neutrino,
$\Gamma (K  \rightarrow \mu N)$ or $\Gamma (\tau \rightarrow \mu \nu N )$, times the branching ratio $Br(N\rightarrow \nu \gamma)$.
This approximation is clearly valid for $N$ with
masses in the range of Eq.~\rf{Range}.  The resulting decay rate formulas are then:
 \begin{figure}[htbp]
\centering
\includegraphics[width=0.75\textwidth,bb=70 650 500 775]{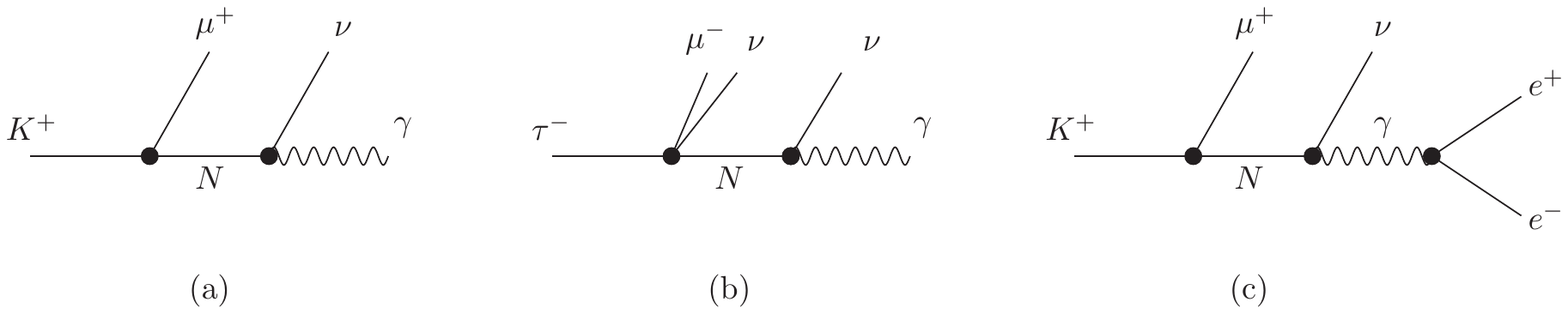}
\caption{Structure of the lowest order contribution of sterile neutrino N  to the radiative decays of K meson (a) and
$\tau$ lepton (b) as well as to the leptonic decay of K meson (c).}
\label{fig-1}
\end{figure}
\begin{eqnarray}\label{DECRATEM}
 \label{DEC-LNV-MUE}
\Gamma^{ }(K^{+}\rightarrow  \mu^+ \nu \gamma) &\approx&
 \Gamma(K^{+}\rightarrow \mu^+ N)\  Br(N \rightarrow \nu \gamma) \\
 \label{DEC-LFV}
 \Gamma^{ } (\tau^{-}\rightarrow \mu^{-} \nu \nu \gamma )&\approx&
\left\{ \Gamma(\tau^{-}\rightarrow \mu^- \nu_\tau N) + \Gamma(\tau^{-}\rightarrow \mu^- \bar \nu_\mu N)\right\} \ Br(N \rightarrow \nu \gamma) ,
 \end{eqnarray}
where the $K$ and $\tau$ decay rates into $N$ are \cite{Shrock}
 \begin{eqnarray}\label{DECAYRATETAU2}
\Gamma(K^{+}\rightarrow \mu^{+} N)&=&|U_{\mu N}|^{2} \frac{G_{F}^{2}}{8\pi} f_{K}^{2} |V_{us}|^{2} m_{K}^{3}
\lambda^{\frac{1}{2}}(x_{\mu}^{2},x_{N}^{2},1)(x_{\mu}^{2}+x_{N}^{2}-(x_{\mu}^{2}-x_{N}^{2})^2) \equiv \
 |U_{\mu N}|^{2} \Gamma_K^{(\mu N)},\\
\label{DEC-TAU-3}
\Gamma(\tau^- \rightarrow \mu^{-}\nu_{\tau} N )&=& |U_{\mu N}|^2
\frac{G_F^2}{192\pi^3} m_\tau^5 I_{1}(z_{N},z_{\nu}, z_{\mu})
\equiv |U_{\mu N}|^2 \Gamma_\tau^{(\mu \nu N)}, \\
\label{DEC-TAU-4}
\Gamma(\tau^- \rightarrow \mu^{-}\bar \nu_{\mu} N )&=& |U_{\tau N}|^2
\frac{G_F^2}{192\pi^3} m_\tau^5 I_{1}(z_{N},z_{\nu}, z_{\mu})
\equiv |U_{\tau N}|^2 \Gamma_\tau^{(\mu \nu N)}.
\end{eqnarray}
Here $f_K = 159$ MeV and  $V_{us}= 0.97377$. We denote $z_{i} = m_{i}/m_{\tau}$, $x_{i} = m_{i}/m_{K}$ with $m_{i} = m_{N}, m_\nu, m_\mu$,  and we use the well known phase space function $\lambda(a,b,c)= a^2+b^2+c^2 -2ab-2bc-2ac$ and the kinematical function $I_1(x,y,z)$ is defined as
\ba{kin-fun-1}
 &&I_{1}(x,y,z)= 12 \int\limits_{(x+y)^{2}}^{(1-z)^{2}} \frac{ds}{s}
(s-x^2-y^{2})(1+z^2-s) \lambda^{1/2}(s, x^{2}, y^2) \lambda^{1/2}(1, s, z^2),
%
 \label{FP-1}
\ea
In the scenario under consideration the decay mode $N\rightarrow \nu\gamma$ is dominant, and therefore as a reasonable approximation,
\begin{eqnarray}\label{dominant}
Br(N\rightarrow \nu\gamma)\approx 1.
\end{eqnarray}

A general issue to take into account in the radiative decays in question is that the intermediate neutrino $N$ propagates as a real particle and decays at a certain distance from the production point. If this distance is larger than the size of the detector, the neutrino $N$ escapes before decaying and the signature of
$\tau\rightarrow \mu \nu \nu \gamma$ or $K \rightarrow  \mu \nu \gamma$  cannot  be recognized. Therefore, in order to calculate the rate of radiative $\tau$ or meson $K$ decays within the detector, one should multiply the theoretical rates (\ref{DECRATEM}) and (\ref{DEC-LFV}) by the probability  $P_{N}$ that the neutrino $N$ decays inside the detector.  Roughly  for a  detector of length $L_{D}$, the probability  $P_{N}$ takes the form
\cite{Atre:2009rg}:
\begin{eqnarray}\label{PN}
P_{N}\approx1-e^{-{L_{D}}/{\tau_N}}
\end{eqnarray}
However, for short enough lifetimes such as \ $\tau_N \lesssim 10^{-9} (s)$ in Eq. \rf{Range},  and detectors of size $L_D \gtrsim 70$ cm, which is typical for this kind of experiments, we can use $P_N \approx 1$.

\begin{figure}[htbp]
\centering
\includegraphics[width=0.75\textwidth]{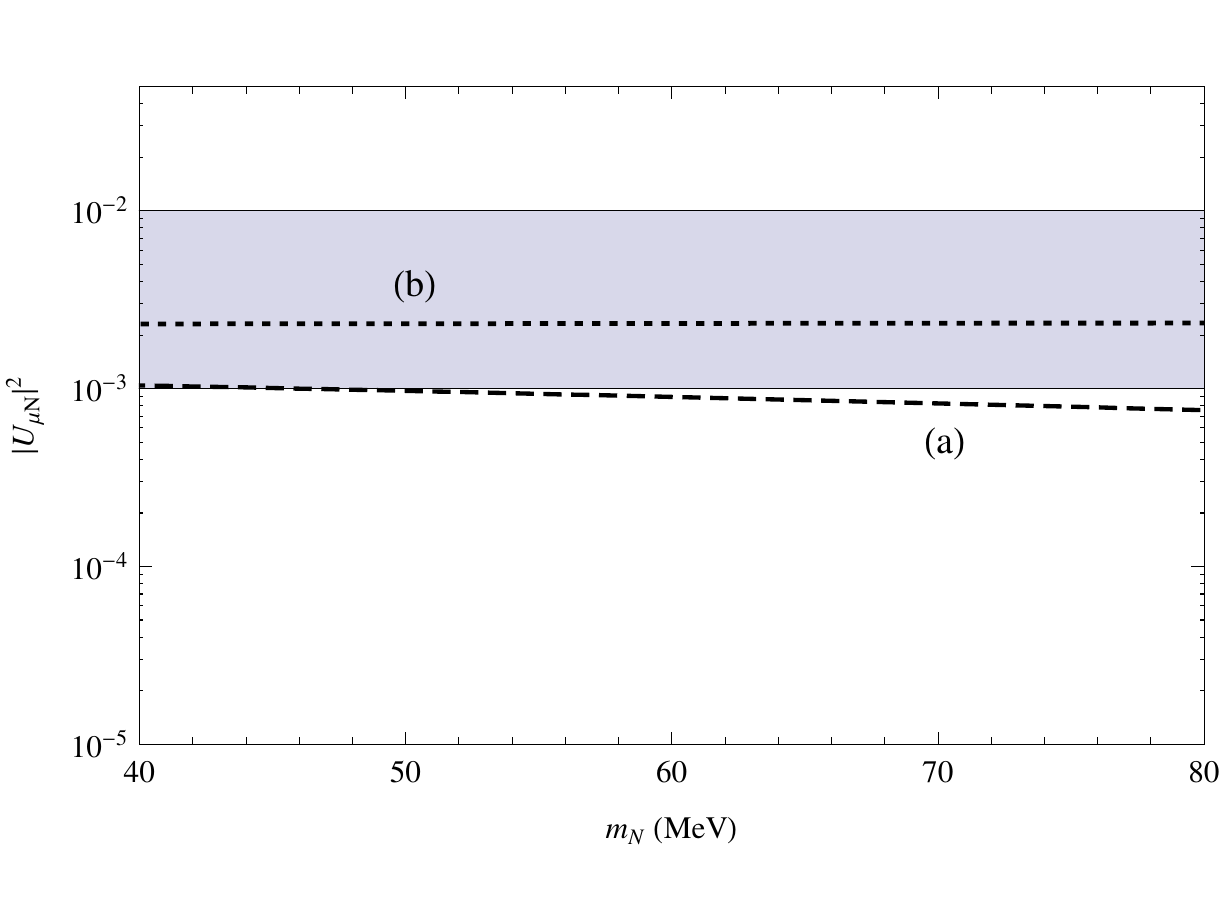}
\caption{The sterile neutrino mass $m_{N}$ and mixing $U_{\mu N}$ with $\nu_{\mu}$. In the gray region
the resolution  \cite{Gninenko,LimitGninenko} of the LSND and MiniBooNE anomalies is possible. The exclusion curves
(a), (b)  are derived from the experimental data (\ref{Data-PL}),  (\ref{Data-PL-2}) respectively. The regions above these curves are excluded.}
\label{fig-2}
\end{figure}

In Ref. \cite{LimitGninenko} the author studied the consistency of a sterile neutrino with parameters in the range given in Eq.\ \rf{Range} with  the data of several experiments, and found no constraints for this part of the parameter space.
Here,  with the same purpose, we examine the following experimental data
\cite{PDG}:
\begin{eqnarray}\label{Data-PL}
Br(K^{+}\rightarrow \mu^{+} \nu \gamma) = (6.2 \pm 0.8) \times 10^{-3}, \\
\label{Data-PL-2}
\ \  \  \ \ \
Br(\tau^{-}\rightarrow \mu^{-} \nu \nu \gamma) = (3.6 \pm 0.4) \times 10^{-3}.
\end{eqnarray}
These measured branching ratios agree with the SM prediction within the quoted experimental uncertainty, namely
 $\Delta^{exp} =$ $0.8 \times 10^{-3}$ and $0.4 \times 10^{-3}$, respectively.
Therefore, the  additional contribution of a sterile neutrino to these processes  should not exceed the respective experimental uncertainties.  Using (\ref{DECRATEM}), (\ref{DEC-LFV}), (\ref{dominant}), (\ref{Data-PL}) and
(\ref{Data-PL-2}) we find the limits
\ba{Limit}
|U_{\mu N}|^2 < \frac{\Delta^{exp}(K^{+}\rightarrow \mu^{+} \nu \gamma)}{\Gamma_{K}^{(\mu N)}/ \Gamma_K} \\ \label{Limit2}
|U_{\mu N}|^2 < \frac{\Delta^{exp}(\tau^{-}\rightarrow \mu^{-} \nu \nu \gamma)}{\Gamma_{\tau}^{(\mu \nu N)}/ \Gamma_\tau}
\ea
valid for a sterile neutrino in the range given in Eq.\ \rf{Range}.
Here $\Gamma_K^{(\mu N)}, \Gamma_\tau^{(\mu \nu N)}$ were defined in (\ref{DECAYRATETAU2}), (\ref{DEC-TAU-3}), (\ref{DEC-TAU-4}). The limits on $|U_{\mu N}|$ given in Eqs.\ \rf{Limit} and (\ref{Limit2}) are plotted in Fig.\ \ref{fig-2}, curves (a) and (b), respectively.
As  shown, The most stringent exclusion curve is Fig.\ 2.a, derived from the $K$ decay data (\ref{Data-PL}). Clearly, this bound is close, but is still  unable to definitely rule out  the whole range of sterile neutrino parameters in Eq.\ \rf{Range} shown in Fig. \ref{fig-2} as the gray zone. On the other hand, the experimental data on radiative $\tau$ decays shown in Eq.\ \ref{Limit2} is consistent with the required parameters.

Nevertheless, the following comment is in order. As we just saw, the experimental measurements of  radiative $K$ decays
are marginally constraining the sterile neutrino parameters of Eq.\  \rf{Range}. However, if experimental cuts were included to restrict the domain of the muon and photon energies, $E_{\mu}$  and $E_{\gamma}$,
characteristic for this mechanism, more stringent bounds can be found.  This is so because in the $K$ rest frame the
muon is monoenergetic with a value of kinetic energy determined by the sterile neutrino mass
\ba{Emu}
E_{\mu(K)} = \frac{(m_K- m_\mu)^2 - m_N^2}{2 m_K}.
\ea
For  $m_N = (40-80)$ MeV as specified in Eq.\ \rf{Range},  the muon energy $E_{\mu(K)}$ varies in a very narrow range \mbox{$E_{\mu(K)} = (146-151)$ MeV.}
In turn, the photon energy in the $K$ rest frame ranges within the interval
\begin{eqnarray}\label{E-gamma}
\frac{1}{2}\left(E_{N} -\sqrt{E^{2}_{N}-m_{N}^{2}}\right)\leq E_{\gamma} \leq
\frac{1}{2}\left(E_{N} + \sqrt{E^{2}_{N}-m_{N}^{2}}\right),
\end{eqnarray}
where $E_{N}$ is the sterile neutrino energy, also a fixed value:
\begin{eqnarray}\label{E_{N}}
E_{N} = \frac{m_{K}^{2} - m_{\mu}^{2} + m_{N}^{2}}{2 m_{K}}.
\end{eqnarray}
For the required range of parameters of Eq.\ \rf{Range}, the photon energy, unlike that of the muon,  is within a rather broad range $E_{\gamma} = (6.8 - 235)$ MeV.

One last test concerns the large value of the neutrino transition magnetic moment, required in the explanation of the LSND and MiniBooNE anomalies. If it exists, this hypothetical parameter must also appear in the process $K^+\to\mu^+\nu e^+e^-$, via a contribution where the photon is virtual and decays into an $e^+ e^-$ pair as shown in the diagram  Fig 1(c).

If the proposed sterile neutrino exists, it can dominate the decay $K^+\to\mu^+\nu e^+e^-$ by an amplitude with the sterile neutrino on mass shell in the intermediate state. The decay rate then factorizes as:
\ba{munuee}
\Gamma(K^+\to\mu^+\nu e^+ e^-)_N = \Gamma(K^+\to\mu^+ N)\times Br(N\to\nu e^+ e^-)
\ea
This representation is valid for sterile neutrino masses within the interval $2m_{e} \leq m_{N} \leq m_{K}-m_{\mu}$.
The first subprocess, $K^+\to\mu^+ N$, can be easily estimated from $K_{\mu2}$, except for a kinematic correction due to the neutrino mass $m_N$ and a factor $|U_{\mu N}|$ due to the $\nu_\mu$ admixture in $N$ (see Eq.\ \ref{DECAYRATETAU2}). The second subprocess is mediated by a photon coupled to the neutrino transition current, which depends on two form factors, $F_1$ and the transition magnetic moment $\mu_{tr}$:
\begin{eqnarray}\label{Nnugamma}
J^\mu _{(N\nu)} = \bar\nu \left\{ F_1 (q^2\gamma^\mu -\not q q^\mu) + i \mu_{tr}\ \sigma^{\mu\nu}q_\nu \right\} N .
\end{eqnarray}
For a real photon, only $\mu_{tr}$ contributes, as in:
\ba{Ntonugamma}
\Gamma(N\to \nu \gamma) = \frac{\mu_{tr}^2 m_N^3}{8\pi} ,
\ea
 while for a virtual photon both $F_1$ and $\mu_{tr}$ contribute without interfering, as in $\Gamma(N\to \nu e^+ e^-)$. Consequently,
\mbox{$\Gamma(N\to \nu e^+ e^-)$} has as lower bound the expression where $F_1$ is neglected, which can be written as:
\ba{Ntonuee}
\Gamma(N\to \nu e^+ e^-) > \frac{8 \alpha_{em}}{3\pi} \left(\log\left(\frac{m_N}{2 m_e}\right)-2/3\right)  \Gamma(N\to \nu \gamma) \sim 10^{-2} \   \Gamma(N\to \nu \gamma).
\ea
Since the experimental measurement \cite{E865}:
\ba{DataKee}
Br(K^{+}\rightarrow \mu^{+} \nu e^+ e^-) = (7.06 \pm 0.31) \times 10^{-8}
\ea
confirms its SM theoretical estimate, then the extra contribution due to the sterile neutrino (see Eq.\ \rf{munuee}) should be at most of the size of the quoted error, thus imposing the bound:
\ba{Kmunueebound}
Br(K^{+}\rightarrow \mu^{+} \nu e^+ e^-)_N < 0.31 \times 10^{-8}.
\ea
Eqs.\  \rf{munuee}, \rf{Ntonuee} and \rf{Kmunueebound} would then impose the bound $Br(K^+\to\mu^+ N) \times Br(N\to\nu\gamma) < 3.\times 10^{-7}$.
Recalling  Eq.~(\ref{DECAYRATETAU2}), $Br(K^+\to\mu^+ N) =  |U_{\mu N}|^{2} \Gamma_K^{(\mu N)}/\Gamma_K$ $\gsim 0.6\times |U_{\mu N}|^{2}$,  we could draw the following bound:
\ba{Thebound}
 | U_{\mu N}|^{2} \times Br(N\to\nu\gamma) < 0.5\times 10^{-6}.
\ea

This stringent bound would rule out the explanation of the LSND and MiniBooNE anomalies in terms of a sterile neutrino with large transition magnetic moment. However, it is not applicable for $m_N$ in the required range of Eq.~\rf{Range}, because the experimental result, Eq.~\rf{DataKee}, is obtained using a cut on the invariant mass of the $e^{+}e^{-}$ pair \mbox{$m_{ee}> 145$ MeV} \cite{E865}. On the other hand, the limit in Eq.~\rf{Thebound} shows that a new measurement of $K^{+}\rightarrow \mu^{+} \nu e^+ e^-$ would provide a stringent bound on the sterile neutrino hypothesis if the cut on $m_{ee}$ could be reduced below 40 MeV.
This 145 MeV cut in the work of Ref.~\cite{E865} had to be applied  in order to suppress the background from the sequence of decays $K^{+}\rightarrow \mu^{+} \nu \pi^{0}$, $\pi^{0}\rightarrow \gamma e^{+}  e^{-}$\cite{Poblaguev}. Then an improvement in the efficiency of the veto for the photons from
$\pi^{0}$-decay and  measurements of the kaon tracks for better control of the missing mass may be required to achieve this goal.


\bigskip

{\it In conclusion.} We have shown that  the existence of a sterile neutrino with mass and mixing in the range given in Eq.\ \rf{Range} is in tension with the existing experimental data on the radiative $K$ meson decay rate, given in Eq.~(\ref{Data-PL}). Future measurements of this rate with better precision will probably be able to derive a more decisive conclusion on the studied question.   In addition, the purely leptonic 4-body $K$ decay $K^{+}\rightarrow \mu^{+} \nu e^+ e^-$ will be able to probe the parameter region of Eq.~\rf{Range} required for the explanation of the LSND and MiniBooNE anomalies, if future measurements reduce the cut in the invariant mass of the $e^{+}e^{-}$ pair in the final state of this decay below 40 MeV.


\begin{acknowledgments}
We are grateful to William Brooks, Serguei Kuleshov and Marcela Gonzalez for discussions.
We also thank Sergey Gninenko and Andrei Poblaguev for useful comments.
This work was supported by
\mbox{FONDECYT} projects 1100582, 110287, 1070227 and
Centro-Cient\'\i fico-Tecnol\'{o}gico de Valpara\'\i so PBCT ACT-028. C.D. acknowledges partial support from Research Ring ACT119, Conicyt.
\end{acknowledgments}

\end{document}